\documentclass[letterpaper, 10 pt, conference]{ieeeconf}  

\IEEEoverridecommandlockouts                              

\usepackage{graphicx}
\usepackage{subfigure}
\usepackage{amsmath,amssymb,amsfonts}
\usepackage{bm}
\usepackage{algorithm} 
\usepackage{algorithmic}                                    

\usepackage[dvipsnames]{xcolor}
\usepackage{dsfont}
\newcommand{\R}{\mathds{R}} 
\newcommand{\T}{^\mathsf{T}} 
\newcommand{\N}{\mathds{N}} 

\title{\LARGE \bf
PI-CoF: A Bilevel Optimization Framework for Solving Active Learning Problems using Physics-Information
}

\author{Liqiu Dong$^{1}$, Marta Zagorowska$^{2}$, Tong Liu$^{1}$, Alex Durkin$^{1}$ and Mehmet Mercang{\"o}z$^{1}$
\thanks{$^{1}$Liqiu Dong, Tong Liu, Alex Durkin and Mehmet Mercang{\"o}z are with Department of Chemical Engineering, Imperial College London
        {\tt\small m.mercangoz@imperial.ac.uk}}%
\thanks{$^{2}$Marta Zagorowska is with Department of Engineering Cybernetics, Norwegian University of Science and Technology. Marta also kindly acknowledges the funding (writing and revisions) from a Marie Curie Horizon Postdoctoral Fellowship project RELIC (grant no 101063948).
        }%
}

\begin{document}

\maketitle
\thispagestyle{empty}
\pagestyle{empty}

\begin{abstract}
Physics informed neural networks (PINNs) have recently been proposed as surrogate models for solving process optimization problems. However, in an active learning setting collecting enough data for reliably training PINNs poses a challenge. This study proposes a broadly applicable method for incorporating physics information into existing machine learning (ML) models of any type. The proposed method - referred to as PI-CoF for Physics-Informed Correction Factors - introduces additive or multiplicative correction factors for pointwise inference, which are identified by solving a regularized unconstrained optimization problem for reconciliation of physics information and ML model predictions. When ML models are used in an optimization context, using the proposed approach translates into a bilevel optimization problem, where the reconciliation problem is solved as an inner problem each time before evaluating the objective and constraint functions of the outer problem. The utility of the proposed approach is demonstrated through a numerical example, emphasizing constraint satisfaction in a safe Bayesian optimization (BO) setting. Furthermore, a simulation study is carried out by using PI-CoF for the real-time optimization of a fuel cell system. The results show reduced fuel consumption and better reference tracking performance when using the proposed PI-CoF approach in comparison to a constrained BO algorithm not using physics information.

\end{abstract}



\section{Introduction}
Physics-informed machine learning (PIML) refers to the combination of physical prior knowledge for the abstraction of natural behaviours, with data-driven models \cite{4}. It has emerged as an effective way to mitigate the shortage of training data and to ensure the physical plausibility of predictive results \cite{6}. From a process optimization perspective, embedding data-driven models into  optimization problems is quickly becoming a state-of-the-art approach \cite{5}. Among various alternatives for surrogate models, such as regression based on polynomials, regression trees, and Gaussian processes (GPs), neural networks are gaining widespread adoption and recently PINNs have been used for optimization as well \cite{1,2}. The attractiveness of PINNs lies in their practicality for integrating physics information. This is achieved by introducing a physics component into the overall loss function during PINN training, facilitating the satisfaction of first-principle equations of interest. In the PINN framework, all inputs and outputs relevant to the system being modelled can be treated simultaneously during training. This simultaneous treatment is a notable advantage, allowing for the seamless incorporation of complex multi-variable relationships into the learning process. However, in an active learning setting collecting enough data for reliably training PINNs poses a challenge.

In many safety-critical industrial systems, critical constraints need to be enforced with high confidence to guarantee process safety during optimization. GP-based models provide attractive properties for surrogate optimization under these conditions, as they enable quantifying uncertainty of predictive results that can be used to probabilistically enforce constraints and the uncertainty can be used to drive exploration during optimization. Owing to this property, BO based on GPs as are widely adopted in process optimization where experimental evaluation is expensive and costly due to deviations from optimal conditions during exploration \cite{3,7,8,21}.

Compared to PINNs, the field of "physics-informed BO" is still under-studied for process systems engineering applications. Tubbs and Mercang{\"o}z \cite{12} proposed the selection of GP mean functions according to domain knowledge. H{\"a}se et al. \cite{10} augmented BO based on kernel density estimation with smooth approximations to categorical distributions. Hanuka et al. \cite{11} utilized a fast approximate model from physical simulations to design the GP model, which is then used for system optimization. In a similar mechanism to PINN training, the studies in \cite{9} and \cite{14} augmented the standard GP with a structured probabilistic model of the expected system's behavior, where the training of the GP and the physics information is considered simultaneously. However, extension of this approach to cover physics information presented over systems of equations with multiple-inputs and multiple-outputs via multi-output Gaussian process regression remains an open research problem.

In contrast to the above works Yang et al. \cite{13} follow an approach to generate low-fidelity data based on physics information and then models the discrepancy between data generated from low-fidelity models and high-fidelity data obtained from actual observations over a spatial domain. This is achieved using parameterized GPs with hyperparameters identified via optimization. For active learning problems this approach of evaluating discrepancies via sampling over the complete search space is not practical but instead it can be possible to carry out pointwise corrections to address physics and ML model discrepancies during the optimization process. Inspired by this idea, in this paper we follow an approach of using correction factors to accommodate ML models and we optimize the values of these correction factors to minimize the discrepancies between ML model predictions and physics information.

The contributions of our paper are:
\begin{itemize}
 	\item We present PI-CoF as a framework in which additive or multiplicative correction factors - utilized for pointwise inference - are determined through solving a regularized unconstrained optimization problem. This process aims to reconcile physics information with ML model predictions;
 	\item We integrate PI-CoF into a real time process optimization framework considering an active learning setting in the form of a bilevel program;
 	\item We demonstrate the proposed framework on a toy numerical example and then on a realistic real time optimization (RTO) case study for the operation of a multi-stack fuel cell system.
 \end{itemize}

The rest of the paper is structured as follows. Section 2 presents the learning based approach for solving RTO problems based on costrained BO optimization and introduces the need for incorporating physics information into the problem setup. Section 3 presents the PI-CoF framework in the form of a bilvel optimization program. Section 4 provides a simple numerical example to illustrate the proposed method. Section 5 presents a case study showing PI-CoF implemented in an RTO setting as part of a control system architecture for the operation of a fuel cell plant with multiple parallel stacks. Section 6 concludes the paper and indicates potential directions for future work.

\section{Background}
\subsection{Real-Time optimization with Gaussian processes}
We want to find an optimal decision $x$ that satisfies a set of constraints:
\begin{equation}
    \begin{aligned}
        \min_x f(t,x,p(x))\\
        \text{s.t. } g(t,x,p(x))\leq 0\\
        \label{eq:Intro}
    \end{aligned}
\end{equation}
where $f,g:\N\times\R^n\times \R^m\rightarrow \R$, $p:\R^n\rightarrow  \R^m$, $x\in\R^n$, $t\in\N$, and the functional form of $f$ and $g$ is known. The functional form of $p$ is unknown, but we have access to measurements $\hat{p}$. Given changing targets such as changing prices or costs of the RTO problem over time $f$ and $g$ will vary in a known fashion. This formulation is a very natural way to incorporate preexisting knowledge via gray box functions $f$ and $g$ \cite{16} and has been studied for process systems engineering applications as well \cite{17,18}. In this setting, we have unknown relationships between the decision variables $x$ (inputs) and intermediate  (outputs) defined by $p$. We use measurements from the environment (plant) to approximate $p$ as a Gaussian process with mean $\mu(x)$ and standard deviation $\sigma(x)$ \cite{Gaussian_Rasmussen2006}.

To solve \eqref{eq:Intro} using the GP approximation for $p$, Korkmaz et al. \cite{15} introduced a Real-Time Optimization algorithm ARTEO based on safe constrained Bayesian Optimization as:
\begin{equation}
\begin{aligned}
\label{eq:ModificationARTEO}
    \min_{x\in \R^n} f(t,x,\mu(x))-z\sigma(x)\\
    \text{s.t. } g(t,x,\mu(x)+\beta\sigma(x))\leq 0
    \end{aligned}
\end{equation}
with a constant $\beta$ defining the desired confidence level \cite{Safe_Koenig2021}, and $z$ as a design parameter. Using a large positive value for $z$ in the objective function favours solutions with large uncertainty, thus promoting exploration (akin to BO using the Upper Confidence Bound acquisition function). Conversely, setting $z\le 0$ will instead lead to solutions which avoid uncertainty as much as possible, focusing on exploitation. At the same time, this formulation preserves safety thanks to including the confidence in the constraints \cite{15}. 

The problem in \eqref{eq:ModificationARTEO} is solved either periodically or in an event-driven fashion and the optimal solutions are implemented in the environment revealing information in the form of new measurements. The measurements are then used to update the GPs approximating $p$.

\section{Physics-informed optimization with ML models}
Our motivation in this paper is to extend the formulation in \eqref{eq:ModificationARTEO} to incorporate physics information provided in the form of a system of equations to further enable the incorporation of known aspects about the problem for example by providing mass and energy balance equations.

\subsection{Physics-informed optimization}
We extend the problem from \eqref{eq:Intro} as:
    \begin{align}
        \min_x f(t,x,p(x))         \label{eq:PIopt}\\
        \text{s.t. } g(t,x,p(x))\leq 0\\
        F(x,p(x))=0         \label{eq:Dynamics}
    \end{align}
where \eqref{eq:Dynamics} describes the physics of the system with $q$ equations. 

Inserting the approximation $p(x)\approx\mu(x)$ into \eqref{eq:Dynamics} does not guarantee $F(x,\mu(x))=0$ at a pointwise evaluation given the value of the decision variables $x$. Thus we introduce a vector of \emph{correction factors} $c\in\R^m$ such that $\|F(x,\tilde{p}(x,c))\|_2^2$ is minimal. For instance, 
    \begin{equation}
        \tilde{p}(x,c)=\mu(x)+c
         \label{eq:CoFs}
    \end{equation}
represents additive correction factors. An alternative formulation could use the correction factors as a scaling term in a multiplicative form as
    \begin{equation}
        \tilde{p}(x,c)=c\mu(x)
         \label{eq:CoFsMul}
    \end{equation}
    
\subsection{Bilevel optimization}
We now formulate \eqref{eq:ModificationARTEO} and \eqref{eq:PIopt} as a bilevel optimization problem of the form:
    \begin{align}
        \min_{x}\;f(t,x,\tilde{p}(x,c^*))-z\sigma(x)     \label{eq:BilevelMod}
\\
        \text{s.t. } g(t,x,\tilde{p}(x,c^*))+\beta\sigma(x))\leq 0 \label{eq:BilevelCstr}\\
        c^*\in C(x)
    \end{align}
where $C(x)$ is a set of solutions of the reconciliation problem parametrised by $x$:
\begin{equation}
    \min_{c\in\R^m} h(x,c)
    \label{eq:hObjmin}
\end{equation}
with 
\begin{equation}
h(x,c)=\sum_{i=1}^q F^2_i(x,\tilde{p}(x,c))+w^{\T}\|c\|_2^2
\label{eq:hObj}
\end{equation}
where $w\in\R^m$ is a vector of constant weights dependent on the corresponding $\sigma(x)$ of the GP for \eqref{eq:CoFs}. We assume that all variables in \eqref{eq:BilevelMod} are normalized, allowing the use of $\sigma(x)$ values in determining $w$, since the key idea is to assign larger correction factors to predicted outputs with a high uncertainty and smaller correction factors to those with low uncertainty. However, $w$ can also be used as a design parameter to direct corrections towards certain outputs.

The optimization problem \eqref{eq:hObjmin} is an unconstrained problem, but it may be nonconvex due to the form of the system of equations describing the physics \eqref{eq:Dynamics}. To facilitate the solution of \eqref{eq:hObjmin}, we introduce the term $w^{\T}\|c\|^2_2$ as a regularization factor. We expect that the number of outputs represented by $p$ ($m$) and the number of corresponding correction factors are greater than the number of physics equations ($q$), which necessitates a regularized treatment of the inner optimization problem.

It should be noted that the calculation of $g$ at a point $x$ in \eqref{eq:BilevelCstr} utilizes the corrected values rather than the original GPs. In the case of composite functions, the propagation of uncertainty from the inner GPs to the outer functions with known form requires special treatment, e.g. using Monte Carlo sampling approaches. The effect of the correction factors on the probability of satisfaction of $g$ should be treated in a similar way but is beyond the scope of our current work.

We cast the reconciliation problem of GP models and the physics information in the context of an optimization problem but the solution of \eqref{eq:hObjmin} can be seen as part of a physics information reconciled inference scheme at a given $x$ for any kind of supervised ML model used for regression purposes, with the mean $\mu$ and standard deviation $\sigma$.

\section{Illustrative example}

\begin{figure}[t]
    \centering
    \includegraphics[width=0.96\linewidth]{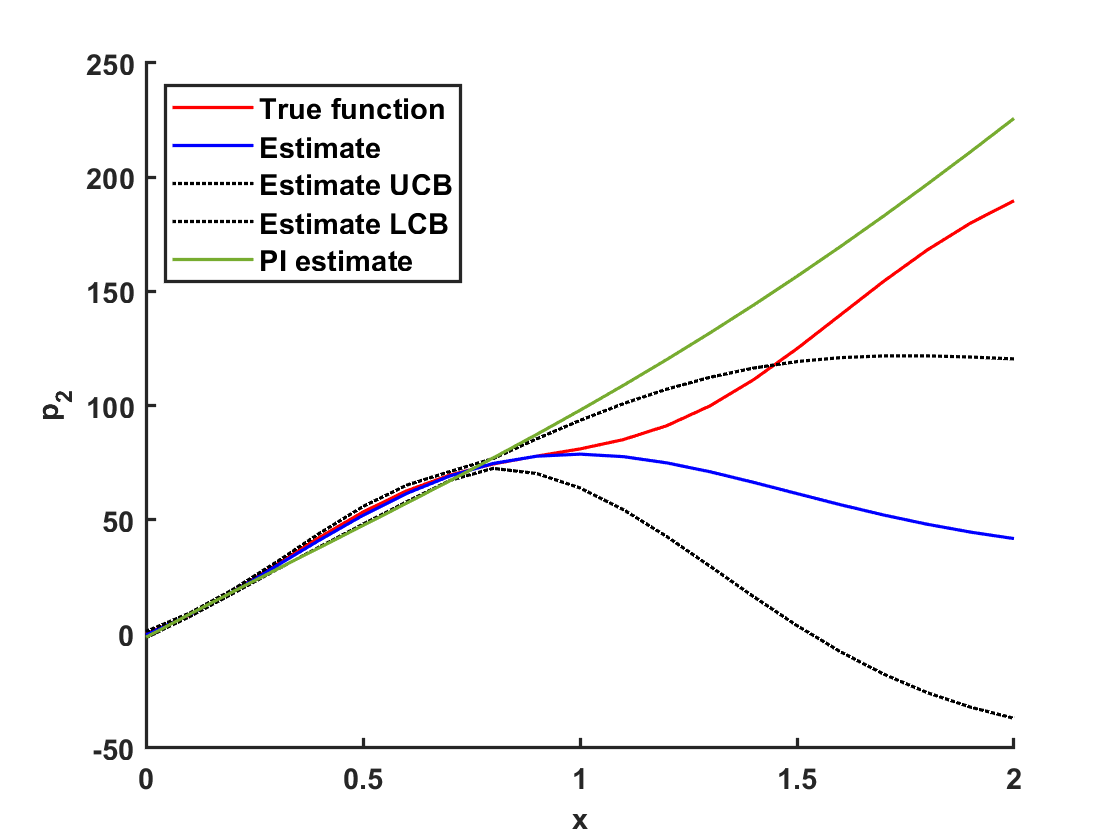}
    \caption{PI-CoF corrected predictions for $p_2$ of the numerical example over the allowed range of the decision variable $x$ at the start of the algorithm.}
    \label{fig:Y1Plot.png}
\end{figure}

We demonstrate the PI-CoF approach first with a numerical example for a system with two outputs and one input, where the true relationship between the inputs and outputs are given with the equations:
\begin{align}
    &{} 120 \sin(0.6x) + 10 \cos(5x) - 10 - p_1 = 0 \label{eq:y_1}\\
    &{} 140 x - p_1 - p_2 = 0  \label{eq:y_2}\
\end{align}
We assume that the form of \eqref{eq:y_1} is unknown but we know \eqref{eq:y_2}, which will serve as the physics information $F$ in the PI-CoF formulation \eqref{eq:BilevelMod}.
We want to solve the following optimization problem concerning the above system:
    \begin{align}
        \min_{x}\;&{}-p_1(x)\\
        \text{s.t. }&{} p_2(x)-140\
    \end{align}
The explicit functional forms of $p_1$ and $p_2$ are approximated with corresponding GP$_i$ with $\mu_i(x)$, $\sigma_i(x)$, $i=1,2$. Using the additive correction factors $c=[c_1,c_2]^{\T}$ as in \eqref{eq:CoFs}, we get from \eqref{eq:BilevelMod} and \eqref{eq:BilevelCstr}:
    \begin{align}
        \min_{x}\;&{}-(\mu_1(x)+c_1^*)-z\sigma_1(x)\\
        \text{s.t. } &{}\mu_2(x)+c_2^*+\beta\sigma_2(x)-140\leq 0 \\
        &{}c^*\in C(x)\label{eq:ReconExample}
    \end{align}
The single equality constraint \eqref{eq:y_2} in the reconciliation problem \eqref{eq:ReconExample} corresponds to the physics and thus $q=1$ in \eqref{eq:hObj} yielding:
\begin{equation}
    h(x,c)=(\mu_2(x)+c_2-140x+\mu_1(x)+c_1)+w^{\T}\|c\|_2^2
\end{equation}
with $w\in\R^2$. We initialize the solution with three samples for $x$ at 0.2, 0.1, 0.05 and the corresponding true values of $p_1$ and $p_2$ to pre-train the two GPs and start the algorithm with a $z$ value of 10 and set $w$ as [0.5\(\sigma_1\), 0.008\(\sigma_2\)]. At the first instance of the algorithm - for illustration purposes - we infer the corrected predictions for $p_2$ over a grid spanning the allowed range 0 to 2 of $x$ by separately solving only the inner problem of PI-CoF and show the result in Fig. \ref{fig:Y1Plot.png} along with the true value of $p_2$ as well as the mean GP prediction and the 95\% confidence interval.

\begin{figure}[htbp]
\begin{center}
\includegraphics[width=0.47\textwidth]{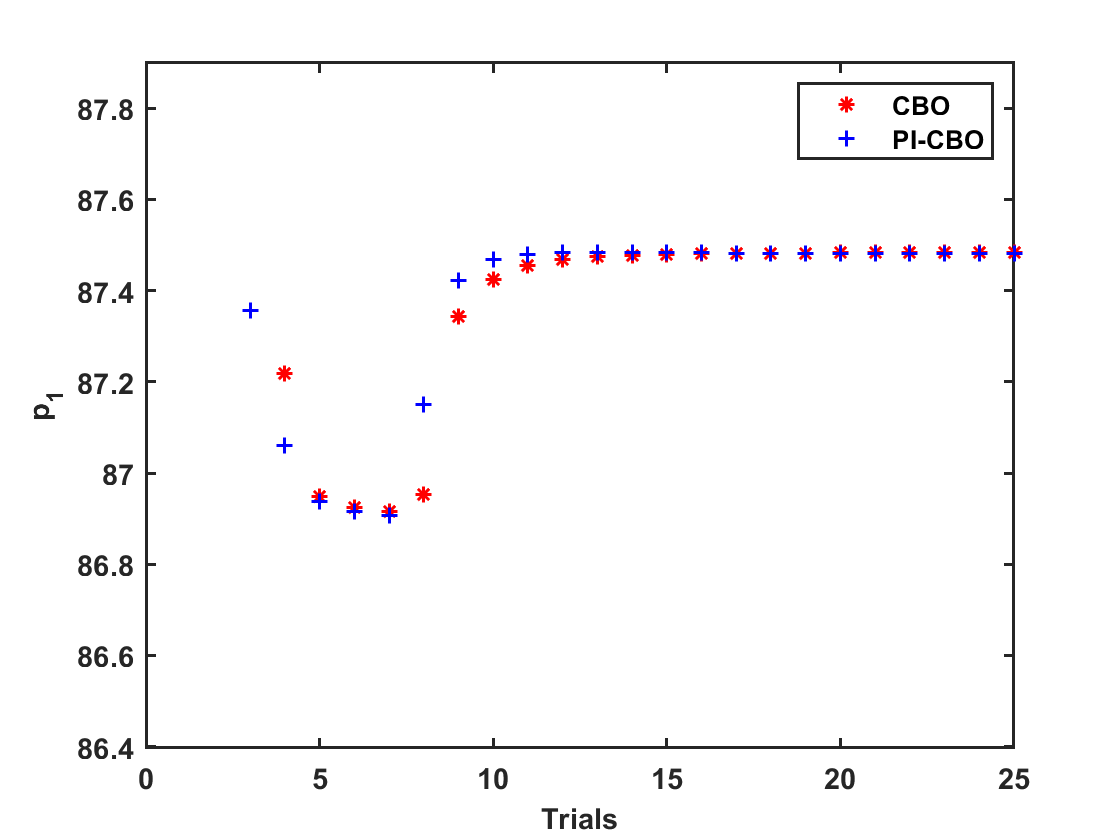}    
\caption{The evolution of the optimization objective for maximizing $p_1$ with PI-CoF and constrained BO} 
\label{fig:3OBJPlot.png}
\end{center}
\end{figure}

\begin{figure}[htbp]
\begin{center}
\includegraphics[width=0.47\textwidth]{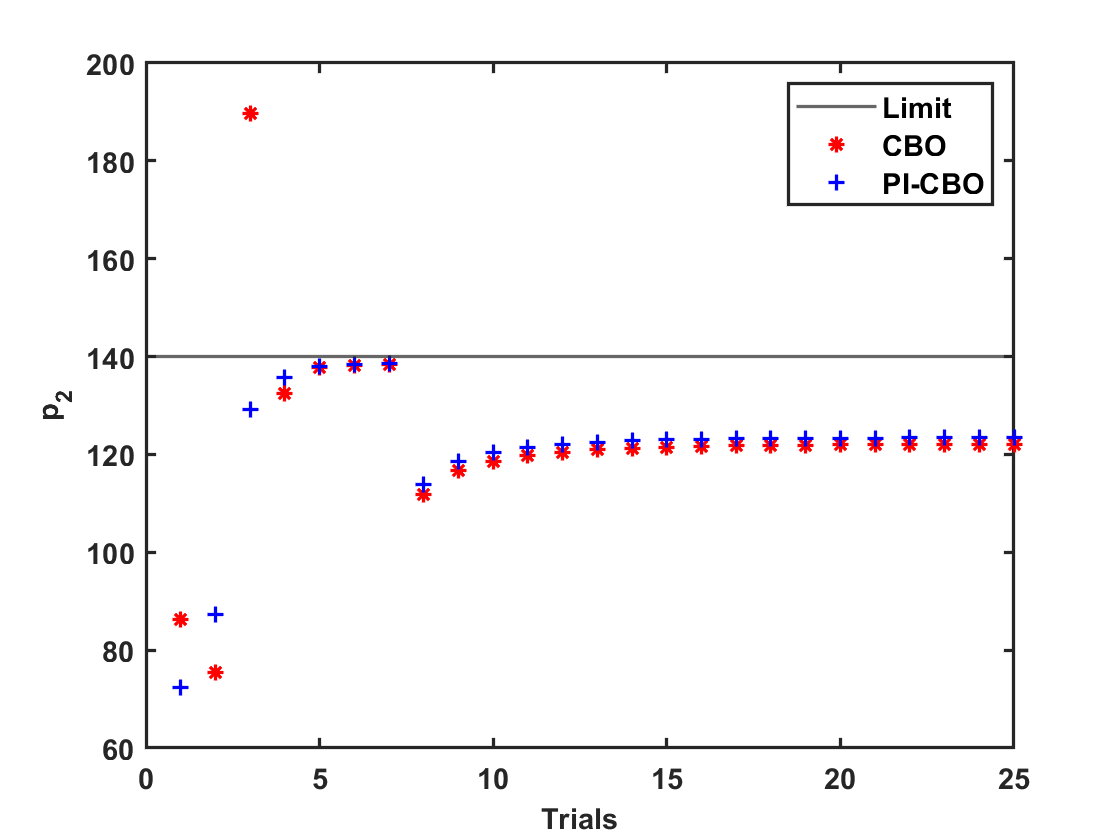}  
\caption{The value of $p_2$ obtained throughout the BO trials and the limit set at 140. Results shown for PI-CoF and constrained BO} 
\label{fig:LimPlot}
\end{center}
\end{figure}

Figure \ref{fig:Y1Plot.png} demonstrates that the PI-CoF estimate is consistently overestimating the value of $p_2$. This is due to the low weight purposefully selected to generate a strong correction for this output, since it influences the satisfaction of $g$. We show the outcomes of running PI-CoF for 25 trials in Fig. \ref{fig:3OBJPlot.png} and Fig. \ref{fig:LimPlot}. The results indicate that compared to a constrained BO formulation shown in \eqref{eq:PIopt}, which is not using the PI-CoF corrections, PI-CoF attains similar or better optimal solutions. Furthermore, as opposed to the constrained BO solution, which leads to a constraint violation, the PI-CoF solution satisfies the constraint throughout the trials. 

\begin{figure}[b]
\begin{center}
\includegraphics[width=0.47\textwidth]{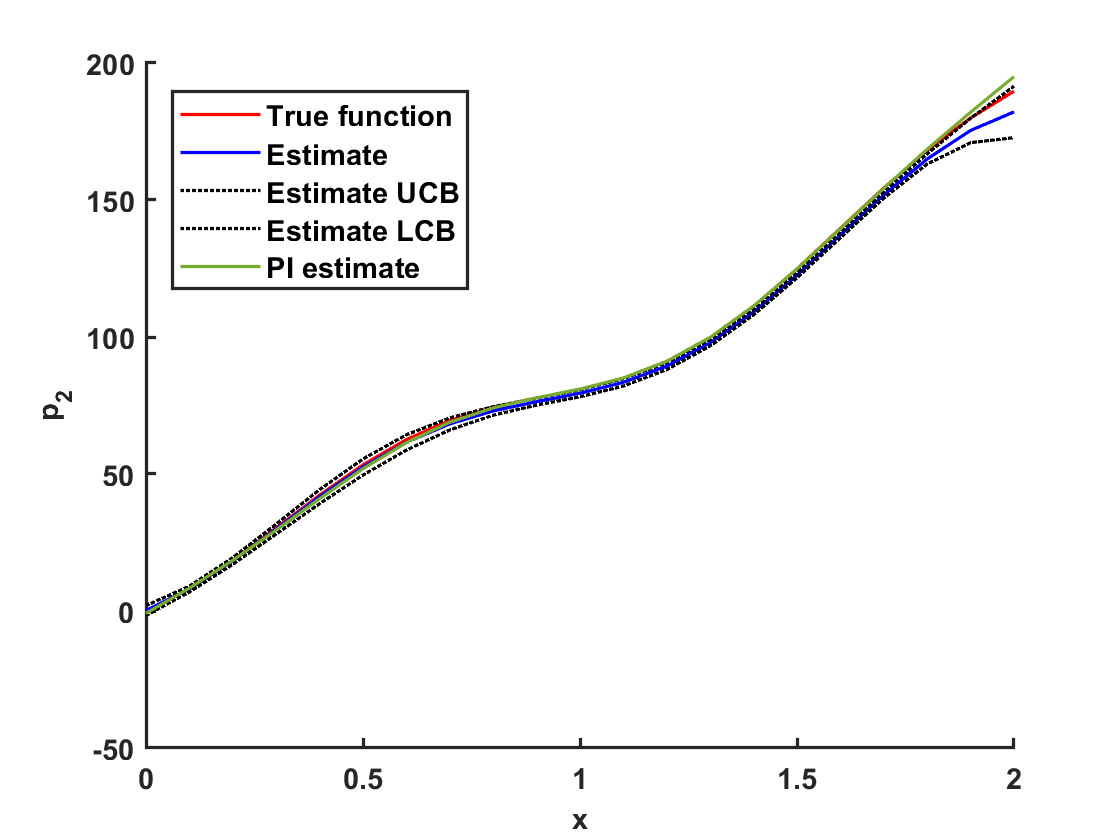}    
\caption{PI-CoF corrected predictions for $p_2$ of the numerical example over the allowed range of the decision variable $x$ at the end of the algorithm} 
\label{fig:Y2Plot_e}
\end{center}
\end{figure}

After the final trial of the PI-CoF algorithm, we again solve the inner problem of PI-CoF over $x$ and plot the result in Fig. \ref{fig:Y2Plot_e} along with the true value of $p_2$ as well as the mean GP prediction and the 95\% confidence interval. It can be seen that the highest values of $p_2$ are never explored as they fall beyond the set constraint limit and PI-CoF is still providing an overestimation in this unexplored region.

\section{Case study: Optimization of Parallel Fuel Cell Stacks}

\begin{figure}[t]
\begin{center}
\includegraphics[width=0.47\textwidth]{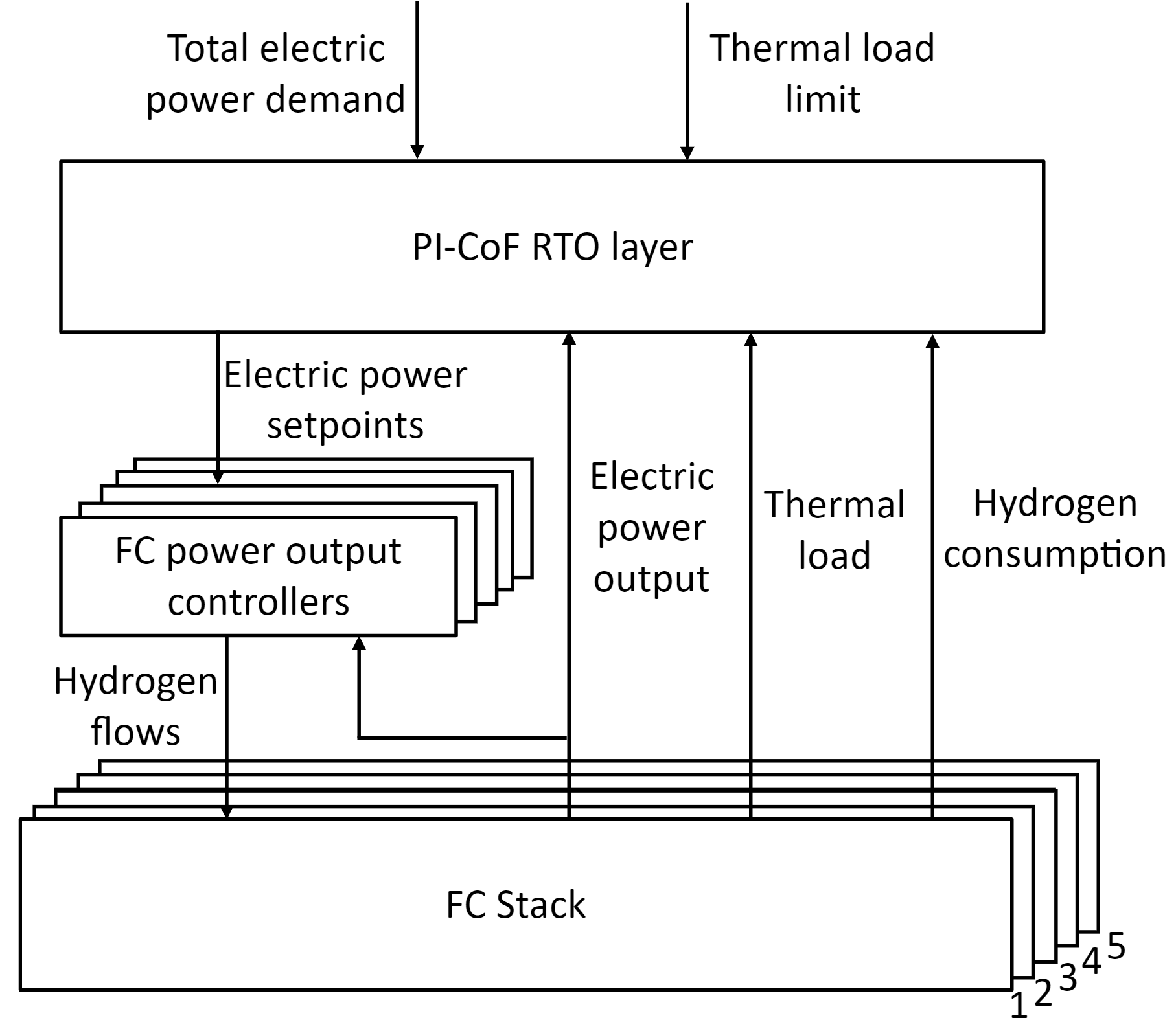} 
\caption{Control system architecture for the case study problem} 
\label{fig:FC Setup}
\end{center}
\end{figure}

We next implement the proposed PI-CoF approach in an RTO setting for the simulated operation of a combined heat and power system composed of multiple parallel stacks of fuel cells (FCs) using hydrogen as a fuel. The FC operating characteristics are adopted from \cite{19}. There are five FC units in operation each with their own power control system. The FCs exhibit different performance characteristics for electric power generation and thermal load given a hydrogen flow rate, which are assumed to be unknown to the optimizer in this case study. The system is sent a total electric power output reference to follow and there is a total thermal load limit to be respected based on the sum of the individual thermal loads of the different stacks. The objective of the optimizer is to satisfy the desired total electric power output, while minimizing fuel consumption and staying below the thermal load limit constraint. The control system architecture for the case study is shown in Fig. \ref{fig:FC Setup}, which is implemented in MATLAB/Simulink with the underlying dynamic equations and controllers for the simulation of the FCs.

The degrees of freedom for the RTO problem are the individual electric power output setpoints provided to the FC power controllers. Therefore, for implementing the RTO problem in the PI-CoF framework, we take the electric power output as the input variable $x$ and the resulting hydrogen consumption and the thermal load as outputs $p$ for each cell. The optimization problem we want to solve is:

\begin{align}
    \min_{x}\;&{} \alpha_1 \sum_{i=1}^{5} p_{hi}^2(x_{it}) + \alpha_2 \left( E_t - \sum_{i=1}^{5} x_{it} \right)^2\\
    \text{s.t. }&{} \sum_{i=1}^{5} p_{thi}(x_{it}) - T_{lim}\leq 0
    \end{align}
where $T_{lim}$ indicates the thermal load limit to be respected and $E_t$ is the reference for total electrical power output. $\alpha_1$ and $\alpha_2$ are the weighting factors to determine the balance between satisfying the desired total electrical power output and minimizing hydrogen consumption. 
The physics information utilized in this case study is the satisfaction of the energy balance based on the combined enthalpy $\Delta H_{\text{rxn}}$ of the reactions taking place in the FCs:
\begin{equation}
F_i = p_{hi}(x_{it})\Delta H_{\text{rxn}} - p_{thi}(x_{it}) - x_i
\label{eq:Physicscase}
\end{equation}

\begin{figure}[h]
\begin{center}
\includegraphics[width=0.47\textwidth]{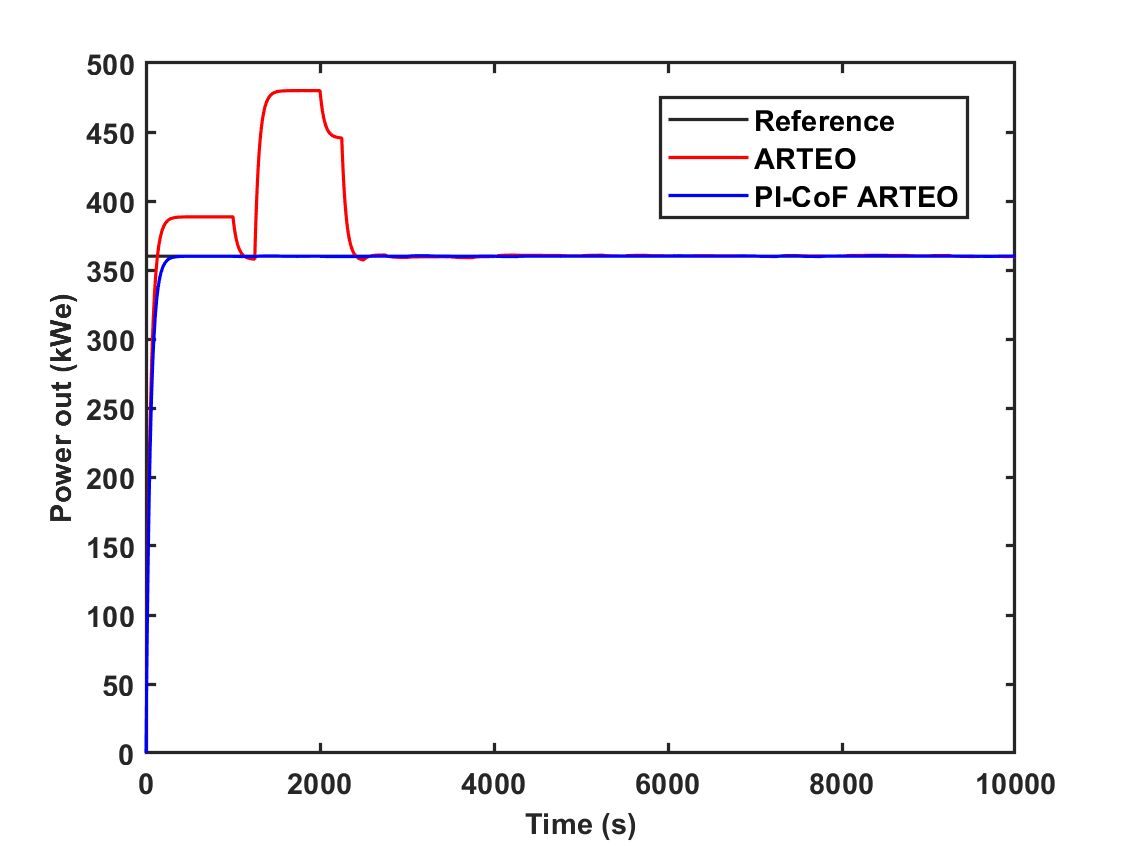} 
\caption{Total power output trajectory and the power output reference} 
\label{fig:PowerOut}
\end{center}
\end{figure}
\begin{figure}[h]
\begin{center}
\includegraphics[width=0.47\textwidth]{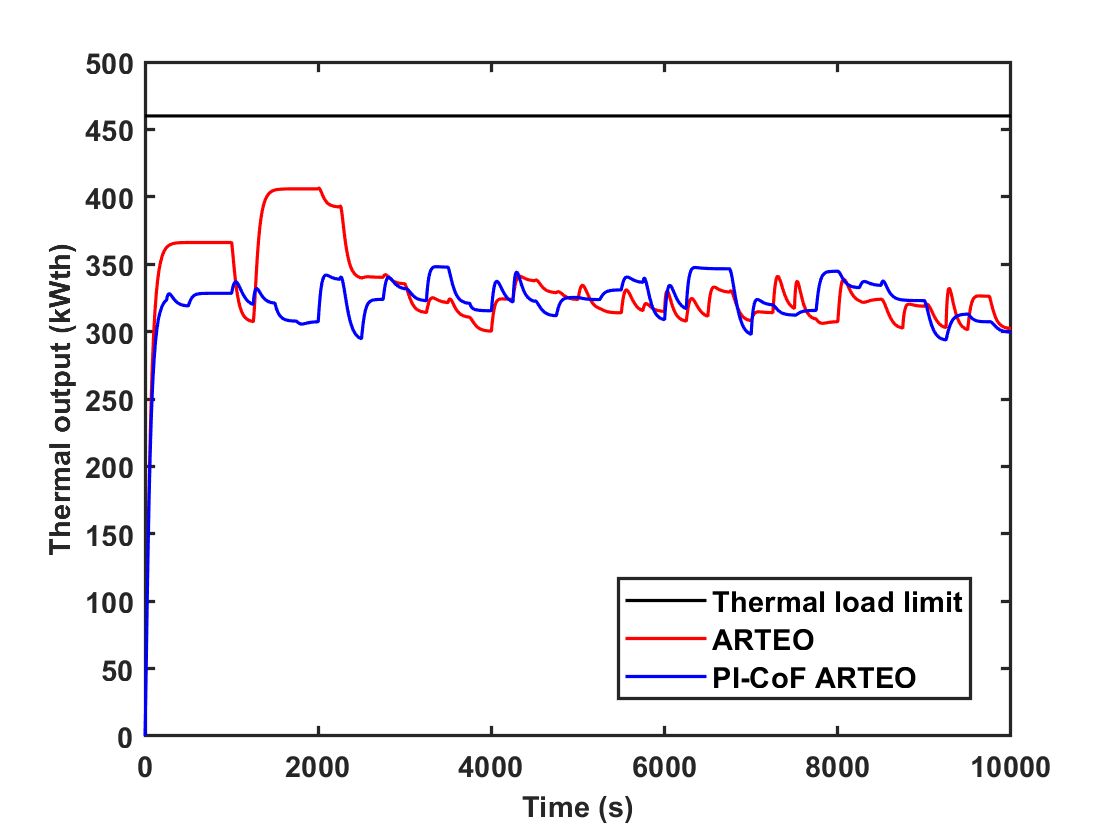} 
\caption{Total thermal load trajectory and the thermal load limit} 
\label{fig:ThermalOut}
\end{center}
\end{figure}
\begin{figure}[h]
\begin{center}
\includegraphics[width=0.47\textwidth]{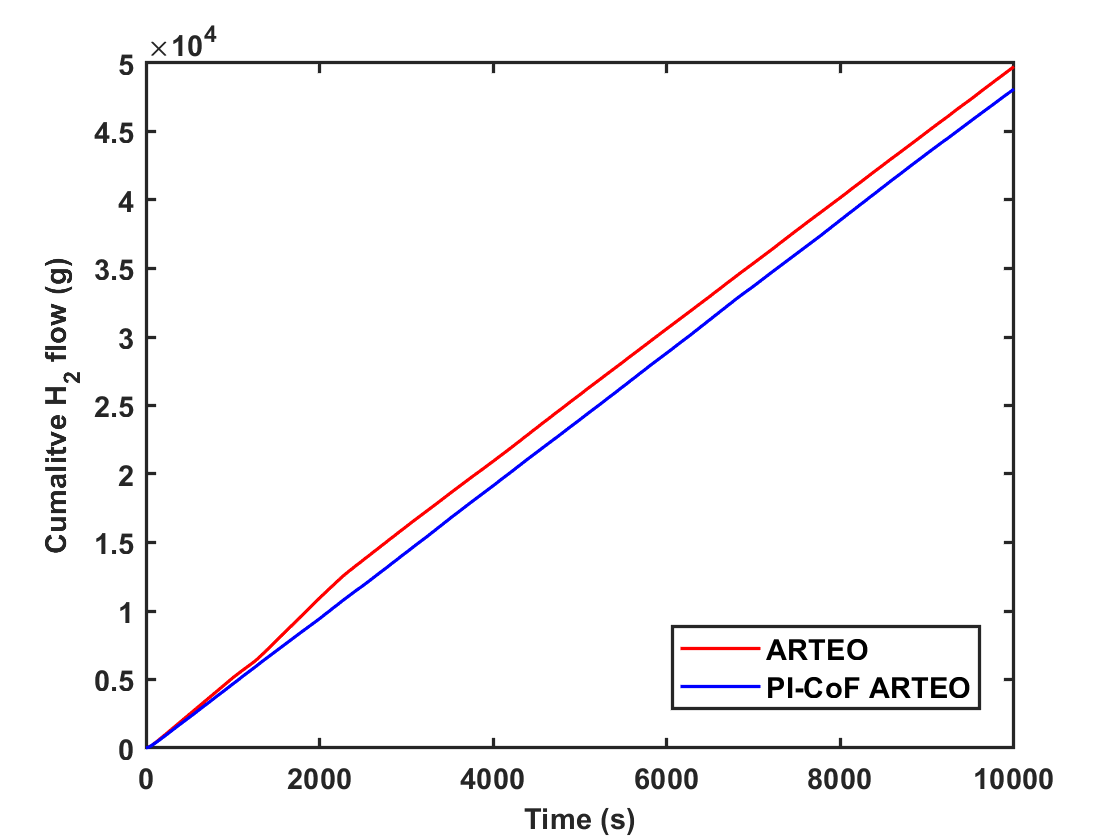} 
\caption{Total hydrogen consumption trajectory} 
\label{fig:H2Out}
\end{center}
\end{figure}

Since the relationships input-output relationships in $p_hi$ and $p_{thi}$ are unknown, we start by training initial GPs from past measurements to initialize the PI-CoF solution. Important to note is that the data represents measurements of achieved actual electric power outputs rather than the setpoints for physical consistency. The objective function $f$ and constraint function $g$ in PI-CoF are given as:
\begin{align}
    f =&{} \alpha_1 \sum_{i=1}^{5} \left(\mu_{hi}(x_{it}) + c^*_{hi}\right)^2 + \alpha_2 \left( E_t - \sum_{i=1}^{5} x_{it} \right)^2\\
    g =&{} \sum_{i=1}^{5} \left(\mu_{thi}(x_{it}) + \sigma_{thi} + c^*_{thi}\right) - T_{lim}
\end{align}
where $\mu(x)$ and $\sigma(x)$ are associated with a GP indicated with the subscripts $h$ and $th$ for hydrogen consumption and thermal load and the additional subscript $i$ indicates a specific FC stack. We also indicate the PI-CoF determined values for correction factors $c^*$ with the same subscripts for the GPs. The physics from \eqref{eq:Physicscase} is:
\begin{equation}
F_i = (\mu_{hi}(x_{it}) + c_{hi})\Delta H_{\text{rxn}} - (\mu_{thi}(x_{it}) + c_{thi}) - x_i
\label{eq:PIcase}
\end{equation}

The algorithm is executed with $z=10000$, the sampling interval of 250 s, and with a constant total power output reference of 360kWe for a duration of 10000 s. For comparison the default ARTEO algorithm based on \eqref{eq:ModificationARTEO} without the PI-CoF corrections is also run with the same settings. 

The results of the simulations are shown in three plots in Fig. \ref{fig:PowerOut}, Fig. \ref{fig:ThermalOut}, and Fig. \ref{fig:H2Out} displaying the evolution of the total power output trajectory, total thermal load trajectory, and the cumulative hydrogen consumption trajectory respectively. The results indicate an improved performance when using the PI-CoF formulation with the PI-CoF version of the algorithm providing a more consistent total power output and resulting in less hydrogen consumption. Whereas both algorithms converge to the same hydrogen consumption rate as indicated by the slope in Fig. \ref{fig:H2Out} while maintaining the same total power output, the PI-CoF version reaches that solution earlier and with minimal exploration. The algorithm without access to physics information carries out additional exploration at the cost of additional fuel consumption. The presented scenarios took around 180 s to simulate the PI-CoF version and around 72 s for the base version of the algorithm indicating a factor of 2.5 increase of the computational effort incurred for solving the inner optimization problem. It is difficult to generalize from this case study with only 5 decision variables for the outer problem and 10 for the inner problem but generally RTO problems are solved over longer periods and the increase in computational load observed in this case study is expected to be tolerable for increased safety and efficiency.

\section{Conclusion and future work}
Motivated by the prospect of using physics information for active learning, we introduced the PI-CoF framework to incorporate physics information into machine learning models within the context of constrained Bayesian Optimization (BO). This framework entails solving an unconstrained inner optimization problem at every evaluation of the objective and constraint functions of an outer optimization problem to determine correction factors for ML models, with the goal of minimizing deviations between ML model predictions and physics information. Experiments with a simple problem are presented to provide insights into how this approach works. A complete case study covering a realistic FC optimization problem is also provided as a proof of concept for the practical implementation of the proposed approach.

Several open points remain for further investigation. The type of bias corrections used in this paper is known to introduce high variance when there is structural mismatch between the models and reality \cite{22}. We did not observe any chattering in simulated experiments but this aspect needs to be investigated considering more case studies. The regularization approach we introduced is powerful in controlling the size of the corrections and is also intuitive when being used along GP models, however there is a possibility of satisfying the physics by using physically implausible corrections. This can be avoided using problem specific insights in selecting weights for the correction factors or by introducing constraints to the inner problem, the latter will however further increase the computational burden introduced by the bilevel optimization formulation. In the constrained BO or more specifically safe BO context, the impact of the correction factors on the predicted uncertainties for constraint satisfaction is also an open problem requiring further investigation. In our implementations, we treated the correction factors as shifting the mean predictions for the GPs and we assumed the uncertainty quantification around the mean to be still valid.  

\bibliography{ifacconf}             
\bibliographystyle{IEEEtran}

\end{document}